\documentclass[conference,a4paper]{IEEEtran}
\usepackage[table]{xcolor}
\usepackage{balance}

\ifCLASSINFOpdf
  \usepackage[pdftex]{graphicx}
\else
  \usepackage[dvips]{graphicx}
\fi
\usepackage{amsmath}
\usepackage{amssymb}
\usepackage{array}
\ifCLASSOPTIONcompsoc
 \usepackage[caption=false,font=normalsize,labelfont=sf,textfont=sf]{subfig}
\else
 \usepackage[caption=false,font=footnotesize]{subfig}
\fi
\DeclareMathOperator{\Tr}{Tr}
\newcommand\numberthis{\addtocounter{equation}{1}\tag{\theequation}} % To number equations

\usepackage[shortlabels]{enumitem} % To have letters in enumerate
\begin{document}
\title{Multistatic OFDM Radar Fusion of MUSIC-based Angle Estimation}

\author{\IEEEauthorblockN{
Martin Willame\IEEEauthorrefmark{1}\IEEEauthorrefmark{2},   % 1st author, 1st affiliations
Hasan Can Yildirim\IEEEauthorrefmark{2},   % 2nd author, 2nd affiliations
Laurent Storrer \IEEEauthorrefmark{2},    % 3rd author, 3rd affiliations
François Horlin\IEEEauthorrefmark{2},      % 4th author, 4th affiliations
Jérôme Louveaux\IEEEauthorrefmark{1},      % 5th author, 5th affiliations
}                                     % ...
%\\
\IEEEauthorblockA{\IEEEauthorrefmark{1}% 1st affiliations
UCLouvain - ICTEAM: Université catholique de Louvain, Louvain-la-Neuve, Belgium 
\IEEEauthorblockA{\IEEEauthorrefmark{2}% 2nd affiliations
ULB - OPERA: Université libre de Bruxelles, Brussels, Belgium}}
\IEEEauthorblockA{\{martin.willame, jerome.louveaux\}@uclouvain.be,\{hasan.can.yildirim, laurent.storrer, francois.horlin\}@ulb.be}}



\maketitle

\begin{abstract}
This study investigates the problem of angle-based localization of multiple targets using a multistatic OFDM radar. Although the maximum likelihood (ML) approach can be employed to merge data from different radar pairs, this method requires a high complexity multi-dimensional search process. The multiple signal classification (MUSIC) algorithm simplifies the complexity to a two-dimensional search, but no framework is derived for combining MUSIC pseudo-spectrums in a multistatic configuration. This paper exploits the relationship between MUSIC and ML estimators to approximate the multidimensional ML parameter estimation with a weighted combination of MUSIC pseudo-spectrum. This enables the computation of a likelihood map on which a peak selection is applied for target detection. In addition to reducing the computational complexity, the proposed method relies only on transmitting the estimated channel covariance matrices of each radar pair to the central processor. A numerical analysis is conducted to assess the benefits of the proposed fusion.
%In recent years, there has been a growing interest in the use of multistatic orthogonal frequency division multiplexing radar configuration. This study investigates the problem of angle-based localization of multiple targets using this radar system. Although the maximum likelihood (ML) approach can be employed to merge data from different radar pairs, the method requires a high complexity multi-dimensional search process. The multiple signal classification (MUSIC) algorithm simplifies the complexity to a two-dimensional search, but no framework is derived for combining MUSIC pseudo-spectrums in a multistatic configuration. This paper exploits the relationship between MUSIC and ML estimators to approximate the multidimensional ML parameter estimation with a weighted combination of MUSIC pseudo-spectrum. This enables the computation of a likelihood map on which a peak selection is applied for target detection. In addition to reducing the computational complexity, the proposed method relies only on transmitting the estimated channel covariance matrix of each radar pair to the central processor, rather than the full raw data observations that classical data-level fusion rules require. A numerical analysis is conducted to assess the benefits of the proposed fusion and to study the impact of system parameters on the localization performance.
\end{abstract}

\vskip0.5\baselineskip

% \fi
% \IEEEpeerreviewmaketitle

\section{Introduction}
The recent development of Wi-Fi sensing has led
to a growing interest in the use of multistatic orthogonal frequency division multiplexing (OFDM) radars \cite{9941042}. A multistatic radar consists of several radar pairs, where the sensing transmitters (STx) and the sensing receivers (SRx) can either be collocated (i.e. a monostatic pair) or separated (i.e. a bistatic pair). Information collected by each radar pair about the targets is sent to a central processor that estimates their positions. Multistatic radars improve target localization by exploiting spatial diversity, but require a data transfer between the SRx and the central processor, as well as the definition of a fusion rule \cite{704504}.

In \cite{201328}, the authors provide an overview of fusion techniques and describe the different levels at which the fusion can occur. At the lowest level, the data fusion framework fuses raw observations to estimate the target location. All the available information is transmitted to the central processor. At a higher level, parametric fusion of the individual radar pairs decisions can be defined. Soft and hard parameter fusion refer to the decision at the central processor relying on a weighted or unweighted combination of the locally estimated parameters, respectively. These parameters may include the range, the angle-of-departure (AoD), the angle-of-arrival (AoA) or directly the target position $(x,y)$. Although data fusion enhances localization by exploiting all available information, it comes at the cost of an increased information transfer to the central processor. Therefore, when designing a system, it is essential to ensure that the fusion method does not necessitate an excessive data transfer.

In this study, we address the localization of $K$ targets using a multistatic radar through the processing of AoD and AoA at each radar pair \cite{6994289}. Parametric fusion in multistatic OFDM radars is a widely studied topic \cite{8918315,Falcone}. However, there exists no data-level fusion method for this problem with low computational complexity and data transfer.

The maximum likelihood (ML) framework presents an optimal solution for deriving a data fusion combination rule. However, the high computational complexity of the ML approach makes it unfeasible. In practice, the brute force estimation requires a $2K$-dimensional search over the $(x,y)$ position of the $K$ targets. The use of a two-dimensional Fourier transform across the transmitting and receiving antenna arrays of each radar pair is a common approximation for the ML solution. However, this approximation yields inaccurate results when the number of antennas is limited. In comparison, subspace-based methods such as the multiple signal classification (MUSIC) algorithm provide better accuracy than the Fourier transform and lower complexity than the ML approach as a two-dimensional search is possible \cite{17564}. However, there is no framework to combine the outputs of the MUSIC algorithm from multiple radar pairs.

\subsection{Major Contributions}
Our contributions can be summarized as follows:
\begin{itemize}
\item We propose a data-level fusion methodology based on the ML framework for joint AoD/AoA-based localization of $K$ targets by a multistatic OFDM radar. The method exploits the relationship between the MUSIC and ML estimators demonstrated in \cite{17564} and some approximations to reduce the complexity of the $2K$-dimensional search of the ML estimator into $K$ two-dimensional problems solved by MUSIC. 
\item The proposed data-level fusion method can be extended to other systems than OFDM multistatic radars as it relies solely on the transmission of the sample covariance matrix of the estimated channel by each radar pair to the central processor.
\item Numerical simulations show the benefits of this method compare to other approaches and investigate the influence of system parameters on the localization performed by a multistatic radar composed of two OFDM radar pairs.
\end{itemize}

\subsection{Notations}
The vectors and matrices are defined as $\mathbf{a}$ and $\mathbf{A}$, respectively. The trace, the transpose and the Hermitian transpose are denoted $\Tr\left\{\mathbf{A}\right\} ,\mathbf{A}^{\mathrm T}$ and $\mathbf{A}^\textup{H}$, respectively. The Moore-Penrose inverse is defined as $\mathbf{A}^{+} =\left(\mathbf{A}^\textup{H} \mathbf{A} \right)^{-1} \mathbf{A}^\textup{H}$. The identity matrix is denoted $\mathbf{I}$. The expectation operator is denoted $\mathbb{E}[\cdot]$ and the Kronecker product is denoted $\otimes$.
\section{System Model} \label{sec:system_model}
In this study, we investigate a multistatic OFDM radar configuration for target localization in the $(x,y)$ plane. The system comprises several STx-SRx radar pairs. The processed data collected by each radar pair is sent to a central processor. The information obtained is then combined to determine the location of the targets. The data to transmit and the combination rule at the central processor are discussed in Section \ref{sec:MLF}.

To express the channel model, we make the following assumptions.
\begin{enumerate}
\item The OFDM symbols transmitted by the different radar pairs are orthogonal (using frequency, time or code division multiple access), enabling each SRx to process the frame transmitted by its paired STx without interference from other STx.
\item The direct signal between the STx and the SRx and the clutter contributions are suppressed from the estimated channel.
\item Only multipath signals with a single reflection on a target have a significant impact on the observed channel model. Signals with multiple reflections are thus neglected.
\end{enumerate}

\begin{figure}
\centering
\includegraphics[width=0.95\columnwidth]{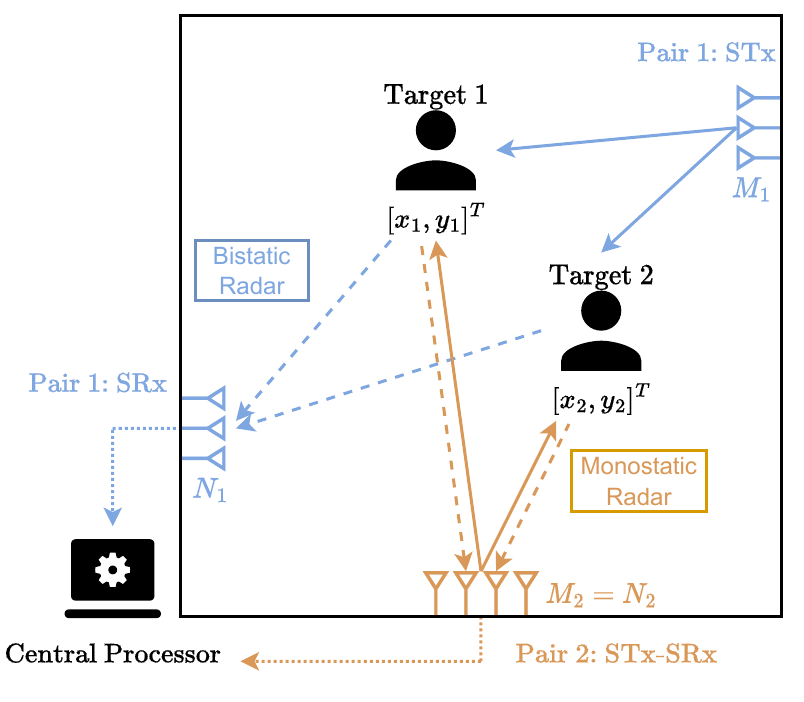}
\caption{Illustration of the scenario. The multistatic configuration is built up of a bistatic radar and a monostatic radar. The solid lines represent the incident waveforms, the dashed lines represent the reflected ones and the dotted lines represent the data transmitted to the central processor.}
\label{fig:scenario}
\end{figure}

The scenario is illustrated in \figurename~\ref{fig:scenario}. The multistatic radar system aims at localizing $K$ targets within its area of coverage. The positions of the $K$ targets are defined by the vectors $\mathbf{x}=[x_1 \ \dots \ x_K]^{\mathrm T}$ and $\mathbf{y}=[y_1 \ \dots \ y_K]^{\mathrm T}$. The system is composed of $P$ radar pairs. Each radar pair consists of an STx and an SRx equipped with a uniform linear array, which for simplicity is assumed  to have half-wavelength spacing and to be oriented toward the coverage area. For the $p\textsuperscript{th}$ radar pair, the AoD and the AoA for the $k\textsuperscript{th}$ target are denoted $\varphi_{p,k}$ and $\vartheta_{p,k}$, respectively. The set of all AoDs and AoAs are denoted by $\boldsymbol \Phi_p$ and $\boldsymbol \Theta_p$. Note that throughout this paper, we simplify notations by omitting the dependence of $(\boldsymbol\Phi_p,\boldsymbol\Theta_p)$ on $(\mathbf{x},\mathbf{y})$. These angles are defined between the waveform and the normal vector of the corresponding antenna array. The $(M_p \times 1)$ AoD and $(N_p \times 1)$ AoA steering vectors are thus given by
\begin{align}
\mathbf{v}_p(\varphi_{p,k})&= [1 \ e^{j\pi \sin(\varphi_{p,k})} \dots \ e^{j\pi (M_p-1) \sin(\varphi_{p,k})}]^{\mathrm T},\\
\mathbf{v}_p(\vartheta_{p,k})&= [1 \ e^{j\pi \sin(\vartheta_{p,k})} \dots \ e^{j\pi (N_p-1) \sin(\vartheta_{p,k})}]^{\mathrm T},
\end{align}
where $M_p$ and $N_p$ denote the number of transmitting and receiving antennas, respectively. The STx of the $p\textsuperscript{th}$ radar pair transmits OFDM symbols isotropically with $Q_p$ subcarriers and a subcarrier spacing of $\Delta_f$. 

The baseband equivalent channel matrix for the $p\textsuperscript{th}$ radar pair is stacked in a vector for each subcarrier $q$ as follows 
\begin{equation} \label{eq:channel}
\mathbf{h}_{p,q}(\boldsymbol\Phi_p,\boldsymbol\Theta_p) = \mathbf{A}_p(\boldsymbol\Phi_p,\boldsymbol\Theta_p) \ \boldsymbol\alpha_{p,q},
\end{equation}
where
\begin{itemize}
    \item $\mathbf{h}_{p,q}$ is the $M_p N_p \times 1$ channel vector of the $p\textsuperscript{th}$ radar pair for subcarrier $q$.
    \item $\mathbf{A}_p(\boldsymbol\Phi_p,\boldsymbol\Theta_p) = [\mathbf{a}_{p,1}(\varphi,\vartheta) \ \dots \ \mathbf{a}_{p,K}(\varphi,\vartheta)]$ is the joint AoD/AoA steering matrix ($M_p N_p \times K$). A short notation $\mathbf{a}_{p,k}(\varphi,\vartheta)$ is used to represent the joint AoD/AoA steering vector defined as $\mathbf{a}_p(\varphi_{p,k},\vartheta_{p,k}) = \mathbf{v}_p(\varphi_{p,k}) \otimes \mathbf{v}_p(\vartheta_{p,k})$.
    \item $\boldsymbol\alpha_{p,q}=[\alpha_{p,q,1} \ \dots \ \alpha_{p,q,K}]^{\mathrm T}$ is the channel coefficient vector ($K \times 1$). These coefficients include the linear increasing phases across subcarriers due to the range of the targets and the attenuation defined by the radar range equation \cite{101049}. Any stochastic model can be associated to the channel coefficient (e.g. the Swerling model \cite{1057561}).
\end{itemize}
\section{Maximum Likelihood Fusion}\label{sec:MLF}
In this section, we develop the ML combination rule for locating the $K$ targets from the information transmitted by every radar pair. The parameters to estimate are defined by the vector $\boldsymbol \gamma = [\mathbf{x}^{\mathrm{T}} \ \mathbf{y}^{\mathrm{T}} \ \{ \boldsymbol\alpha_{p,0}^{\mathrm{T}} \ \dots \ \boldsymbol\alpha_{p,Q_p-1}^{\mathrm{T}} \}_{p=1\dots P}]^{\mathrm{T}}$. The number of targets to localize is assumed to be known as methods to estimate $K$ can be found in the literature \cite{1164557}. At each radar pair, information about the targets is acquired by estimating the channel at the SRx using a known OFDM symbol transmitted by the STx. The observed data from the $p\textsuperscript{th}$ radar pair for the ML development is thus modelled as a noisy channel estimated vector
\begin{equation} \label{eq:observations}
\Tilde{\mathbf{h}}_{p,q} = \mathbf{A}_p(\boldsymbol\Phi_p,\boldsymbol\Theta_p) \ \boldsymbol\alpha_{p,q} + \mathbf{n}_{p,q}, 
\end{equation}
where the $(M_p N_p \times 1)$ vector $\mathbf{n}_{p,q}$ represents the estimation errors which are assumed to be Additive White Gaussian Noise (AWGN) of variance $\sigma^2_p$.

Notice that our study solely focuses on joint AoD/AoA-based ML localization and does not utilize the range information. As a result, each channel coefficient vector $\boldsymbol\alpha_{p,q}$ is independently estimated, since the linear phase increase across subcarriers defined by the range of each target is not exploited. 

Considering independent noise contributions for the estimated channel vector from all radar pairs, the combined likelihood function is obtained as the product of the individual Gaussian density functions. After taking the natural logarithm of the combined likelihood function, the sum of the individual log-likelihood functions has to be maximized. These individual log-likelihood functions are denoted $\mathcal{L}_p(\boldsymbol\gamma)$ and are developed in the subsection below separately for each radar pair as their contributions are independent.

\subsection{Development of the Individual Likelihood Functions} \label{sec:local_ML_function}
By omitting the dependence on $\boldsymbol \gamma$, for notation simplicity, the individual log-likelihood function for the $p\textsuperscript{th}$ radar pair is given by
\begin{equation} \label{eq:ind_log_likelihood}
\mathcal{L}_p =  \frac{-1}{2\sigma_p^2} \sum_{q=0}^{Q_p-1} \lVert \Tilde{\mathbf{h}}_{p,q} - \mathbf{A}_p(\boldsymbol\Phi_p,\boldsymbol\Theta_p) \ \boldsymbol\alpha_{p,q} \rVert^2.
\end{equation}
First, we maximize with respect to the channel coefficients $\boldsymbol\alpha_{p,q}$ to obtain a closed-form expression as a function of $\boldsymbol\Phi_p,\boldsymbol\Theta_p$. After solving the resulting linear least square problem, the ML estimate of the channel coefficient for every subcarrier $q$ is 
\begin{equation} \label{eq:Moore-Penrose}
\Hat{\boldsymbol\alpha}_{p,q}(\boldsymbol\Phi_p,\boldsymbol\Theta_p) = \mathbf{A}_p^{+}(\boldsymbol\Phi_p,\boldsymbol\Theta_p)\ \Tilde{\mathbf{h}}_{p,q}.
\end{equation}
By inserting the estimates \eqref{eq:Moore-Penrose} back into \eqref{eq:ind_log_likelihood} and after some mathematical steps, the individual log-likelihood function can be rewritten as
\begin{equation} \label{eq:trace_log_likelihood}
\mathcal{L}_p = \frac{Q_p}{2\sigma_p^2} \Tr\left\{ \mathbf{A}_p(\boldsymbol\Phi_p,\boldsymbol\Theta_p)\mathbf{A}_p^{+}(\boldsymbol\Phi_p,\boldsymbol\Theta_p) \ \Tilde{\mathbf{R}}_p\right\},
\end{equation}
in which $\Tilde{\mathbf{R}}_p = \frac{1}{Q_p} \sum_{q=0}^{Q_p-1} \Tilde{\mathbf{h}}_{p,q} ~\Tilde{\mathbf{h}}^\textup{H}_{p,q}$ is the sample covariance matrix of the channel vector averaged over the subcarriers.

We search the set of $(x,y)$ positions of the $K$ targets that maximizes the sum of the individual log-likelihood functions. Therefore, the brute-force maximization of the sum of \eqref{eq:trace_log_likelihood} implies solving a $2K$-dimensional problem due to the presence of multiple targets within the coverage area. This high level of computational complexity renders the solution of the exact ML problem impractical. In the next subsection, we establish the connection between the MUSIC algorithm and the ML. This link is then exploited in Subsection \ref{sec:weights} to replace the $2K$-dimensional ML parameter estimation by $K$ two-dimensional estimations based on MUSIC pseudo-spectrum outputs.

\subsection{From ML to MUSIC} \label{sec:ML_to_MUSIC}
In \cite{17564}, the authors investigate the relationship between the MUSIC and the ML estimators for a single radar pair $p$. The authors assert that the MUSIC estimator is a large sample realization of the ML estimator, if and only if the coefficient covariance matrix $\mathbf{S}_p = \frac{1}{Q_p} \sum_{q=0}^{Q_p-1}\mathbb{E}[\boldsymbol \alpha_{p,q} \boldsymbol\alpha_{p,q}^\textup{H}]$ is diagonal. The large sample assumption holds when the number of subcarriers $Q_p$ is sufficiently large to obtain an accurate estimate of the covariance matrix $\mathbf{R}_p$. As the number of subcarriers also plays a direct role on the diagonality of $\mathbf{S}_p$, its impact on the localization performance is discussed in Section \ref{sec:sim_res}.

By adapting the steps of the proof of \cite[Theorem 6.1]{17564}, we can show that maximizing the log-likelihood function given in \eqref{eq:trace_log_likelihood} is equivalent under the assumptions above to maximizing the following expression
\begin{equation} \label{eq:final_log_likelihood}
\mathcal{L}_p = \frac{Q_p}{2\sigma_p^2} \sum_{k=1}^K \underbrace{\mathbf{a}_{p,k}^\textup{H}(\varphi,\vartheta) \ \Tilde{\boldsymbol \Gamma}_p \ \mathbf{a}_{p,k}(\varphi_,\vartheta)}_{(a)} \ \underbrace{\Hat{s}_{p,k}(\boldsymbol\Phi_p,\boldsymbol\Theta_p)}_{(b)} \ ,
\end{equation}
where $\Tilde{\boldsymbol \Gamma}_p=\Tilde{\mathbf{U}}_p\Tilde{\mathbf{U}}_p^\textup{H} = \mathbf{I}-\Tilde{\mathbf{G}}_p\Tilde{\mathbf{G}}_p^\textup{H}$, where $\Tilde{\mathbf{U}}_p$ and $\Tilde{\mathbf{G}}_p$ are the signal and noise subspace matrices, which are obtained from the singular value decomposition of $\Tilde{\mathbf{R}}_p$. The $K$ eigenvectors corresponding to the strongest eigenvalues form the signal subspace $\Tilde{\mathbf{U}}_p$, the remaining vectors form the noise subspace $\Tilde{\mathbf{G}}_p$. The coefficient $\Hat{s}_{p,k}(\boldsymbol\Phi_p,\boldsymbol\Theta_p)$ is the $k\textsuperscript{th}$ diagonal element of the estimated sample coefficient covariance matrix $\Hat{\mathbf{S}}_p(\boldsymbol\Phi_p,\boldsymbol\Theta_p)$ defined as
\begin{align*}
\Hat{\mathbf{S}}_p(\boldsymbol\Phi_p,\boldsymbol\Theta_p)
& = \frac{1}{Q_p} \sum_{q=0}^{Q_p-1} \Hat{\boldsymbol \alpha}_{p,q}(\boldsymbol\Phi_p,\boldsymbol\Theta_p) \ \left(\Hat{\boldsymbol\alpha}_{p,q}(\boldsymbol\Phi_p,\boldsymbol\Theta_p)\right)^\textup{H} \\
& = \mathbf{A}_p^{+}(\boldsymbol\Phi_p,\boldsymbol\Theta_p) \ \Tilde{\mathbf{R}}_p \ \left(\mathbf{A}_p^{+}(\boldsymbol\Phi_p,\boldsymbol\Theta_p)\right)^\textup{H}. \numberthis \label{eq:S_matrix}
\end{align*}
Two parts can be identified in \eqref{eq:final_log_likelihood}:
\begin{enumerate}[(a)]
\item is equivalent to the output of the MUSIC pseudo-spectrum for the tested angles $(\varphi_{p,k},\vartheta_{p,k})$ of the $k\textsuperscript{th}$ target. The steering vector $\mathbf{a}_{p,k}(\varphi,\vartheta)$ is projected onto the sample signal subspace using the projection matrix defined as $\Tilde{\boldsymbol \Gamma}_p$.
\item is the estimated signal power received from the $k\textsuperscript{th}$ target. This term represents a weight applied to the output of the MUSIC algorithm of the $p\textsuperscript{th}$ radar pair.
\end{enumerate}
It is worth noting that unlike the MUSIC output which only depends on $(\varphi_{p,k},\vartheta_{p,k})$, the values of $\Hat{s}_{p,k}(\boldsymbol\Phi_p,\boldsymbol\Theta_p)$ solving the ML function in \eqref{eq:final_log_likelihood} still depend on all angles of all $K$ targets, therefore requiring a $2K$-dimensional parameter search.

\subsection{Estimation of the Weighting Coefficients} \label{sec:weights}
In this subsection, we present an alternative approach to the ML estimation by using an approximated expression of the weighting coefficient $\Hat{s}_{p,k}(\boldsymbol\Phi_p,\boldsymbol\Theta_p)$. This method relies on a local pre-estimation of the set of angles. For each radar pair $p$, we propose to pre-estimate the  AoDs $\boldsymbol \Phi_p$ and AoAs $\boldsymbol \Theta_p$ using the $K$ largest peaks of the MUSIC pseudo-spectrum. Then, instead of evaluating $\Hat{\mathbf{S}}_p(\boldsymbol\Phi_p,\boldsymbol\Theta_p)$ for each angles corresponding to the tested positions $(\mathbf{x},\mathbf{y})$ of all targets, the matrix is evaluated for the pre-estimated angles $(\Hat{\boldsymbol \Phi}_p, \Hat{\boldsymbol \Theta}_p)$. This allows to decouple the $2K$-dimensional parameter search of the ML estimator of the target positions into $K$ two-dimensional estimations solved by weighted MUSIC outputs. 

In practice, the position of the targets are obtained by evaluating the log-likelihood function for every point $(x,y)$ of a two-dimensional search grid. For the $p\textsuperscript{th}$ radar pair, the individual log-likelihood function evaluated for a given tested position $(x,y)$ can thus be rewritten as
\begin{equation} \label{eq:2D_log_likelihood}
\mathcal{L}_p(x,y) = \frac{Q_p}{2\sigma_p^2} \ \mathbf{a}_p^\textup{H}(\varphi,\vartheta) \ \Tilde{\boldsymbol \Gamma}_p \ \mathbf{a}_p(\varphi,\vartheta) \  \Hat{\overline{s}}_{p}(\Hat{\boldsymbol\Phi}_p,\Hat{\boldsymbol\Theta}_p),
\end{equation}
where $(\varphi,\vartheta)$ are the AoD and the AoA corresponding to the tested position $(x,y)$ and $\Hat{\overline{s}}_{p}(\Hat{\boldsymbol\Phi}_p,\Hat{\boldsymbol\Theta}_p) = \sum_{k=1}^{K} \Hat{s}_{p,k}(\Hat{\boldsymbol\Phi}_p,\Hat{\boldsymbol\Theta}_p)$. The approximated individual log-likelihood function is thus given by a two-dimensional estimation obtained from weighted MUSIC outputs.

\subsection{Fusion of Multiple Radar Pairs} \label{sec:fusion}

From the expression of the individual log-likelihood function for each radar pair stated in \eqref{eq:2D_log_likelihood}, the combined log-likelihood function for a position $(x,y)$ of the search grid is obtained as the sum of these individual functions
\begin{equation}\label{eq:combined_likelihood}
\mathcal{L}(x,y) = \sum_{p=1}^{P} \frac{Q_p}{2\sigma_p^2} \ \mathbf{a}_p^\textup{H}(\varphi,\vartheta) \ \Tilde{\boldsymbol \Gamma}_p \ \mathbf{a}_p(\varphi,\vartheta) \  \Hat{\overline{s}}_{p}(\Hat{\boldsymbol\Phi}_p,\Hat{\boldsymbol\Theta}_p).
\end{equation}
Then, the estimate of the $K$ target positions is defined by the $K$ largest peaks obtained from this combined log-likelihood map. The combined log-likelihood function in \eqref{eq:combined_likelihood} represents a weighted combination of the MUSIC pseudo-spectrum outputs of each radar pair. The weights are determined by the number of subcarriers $Q_p$, by the noise variance $\sigma_p^2$ and by the estimated received signal power $\Hat{\overline{s}}_{p}(\boldsymbol\Phi_p,\boldsymbol\Theta_p)$, which is derived from a local pre-estimation of the AoD and AoA of the targets. The two-dimensional weighted MUSIC transform followed by a peak selection has replaced the $2K$-dimensional ML parameter estimation. 

The expression of the combined log-likelihood in \eqref{eq:combined_likelihood} can be evaluated by the central processor only from the sample covariance matrix $\Tilde{\mathbf{R}}_p$ of every radar pair $p$. Therefore, the proposed data fusion algorithm requires each radar pair to send an $(M_p N_p)$ square matrix to the central processor, instead of a $(Q_p \times M_p \times N_p)$ estimated channel tensor.
\section{Simulation Results} \label{sec:sim_res}
\begin{table*}[t]
\renewcommand{\arraystretch}{1.3}
\caption{Comparison of the RMSE in meters for the presented combination methods for various radar pair characteristics. Unless otherwise stated, $M_p=4,\ Q_p=512$ and $\sigma_1^2=\sigma_2^2$.}
\label{tab:RMSE}
\centering
\begin{tabular}{|c|c|c|c|c|c|}
\hline
%First row
\rowcolor[HTML]{C0C0C0} 
  & \textbf{Our Method} & \textbf{Method A} &  \textbf{Method B} &  \textbf{2D-FFT}  & \textbf{Soft Fusion}\\ \hline
%Second row
Same characteristics  & $1.55$ & $1.67$ & $2.61$ & $2.97$  &  $2.04$ \\ \hline
%Third row
\rowcolor[HTML]{E8E8E8} 
$M_1=8=2M_2$ & $1.24$& $1.29$ & $2.15$  & $2.52$ &  $1.64$ \\ \hline
%Fourth row
$\sigma_1^2=0.5\sigma_2^2$ & $1.44$ & $1.57$ & $2.62$ & $2.99$  &  $1.90$  \\ \hline
%fifth row
\rowcolor[HTML]{E8E8E8} 
$M_1=8=2M_2,\sigma_1^2=2\sigma_2^2$ & $1.15$ & $1.19$ & $2.19$  & $2.56$ & $1.51$ \\ \hline
\end{tabular}
\end{table*}

In this section, Monte-Carlo simulations are performed to assess the improvement brought by the proposed fusion method on the accuracy of a multistatic radar system to localize targets in the coverage area. The method proposed in Section \ref{sec:MLF} is compared to the following methods:
\begin{description}
\item[Method A] The MUSIC outputs are combined without a weight depending on the received signal power. In \eqref{eq:combined_likelihood}, we define $\Hat{\overline{s}}_{p}=1$.
\item[Method B] Instead of relying on a pre-estimation of $(\Hat{\boldsymbol \Phi}_p, \Hat{\boldsymbol \Theta}_p)$, we can approximate the signal power by assuming that the targets have orthogonal steering vectors. Using this assumption in the expression of the matrix $\Hat{\mathbf{S}}_{p}$ in \eqref{eq:S_matrix}, it comes down to use \eqref{eq:combined_likelihood} with $\Hat{\overline{s}}_{p}(\varphi,\vartheta) = \mathbf{a}_p^\textup{H}(\varphi,\vartheta) \ \Tilde{\mathbf{R}}_p \ \mathbf{a}_p(\varphi,\vartheta).$ 
%\item[Method B] The received signal power is estimated by assuming that the targets have orthogonal steering vectors. The term $\Hat{\overline{s}}_{p}$ in \eqref{eq:combined_likelihood} is then defined as $\Hat{\overline{s}}_{p}(x,y) = \mathbf{a}_p^\textup{H}(\varphi,\vartheta) \ \Tilde{\mathbf{R}}_p \ \mathbf{a}_p(\varphi,\vartheta).$
\item[2D-FFT] This is equivalent to the classical two-dimensional Fourier transform processing. It can be shown to be the optimal ML solution if the targets have orthogonal steering vectors. The combination rule is defined by using \eqref{eq:combined_likelihood} with $\Hat{\overline{s}}_{p}=1$ and $\Tilde{\boldsymbol \Gamma}_p=\Tilde{\mathbf{R}}_p$. 
\item[Soft Fusion] This parameter fusion method sets the detected position of each target to the weighted average of the local decisions of each radar pair \cite{201328}. The weight is based on the local log-likelihood function value with $\Hat{\overline{s}}_{p}=1$. Note that this method requires an association of the detected positions of each radar pair which is here assumed to be perfect.
\end{description}

\begin{figure}
\centering
\includegraphics[trim=3.4cm 9.5cm 4.3cm 9.91cm,width=0.95\columnwidth]{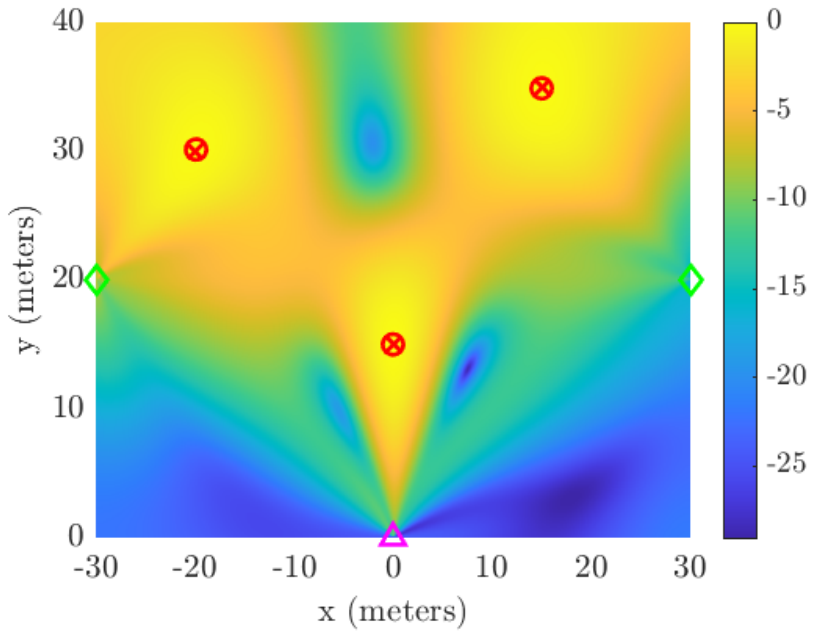}
\caption{Illustration of the combined log-likelihood map in dB, normalized to the value of the highest peak. The two STx are represented by the diamonds, the SRx by the triangle, the true target positions by the crosses and the estimated location of the targets by the circles.}
\label{fig:setup}
\end{figure}

\begin{figure}
\centering
\includegraphics[width=0.95 \columnwidth]{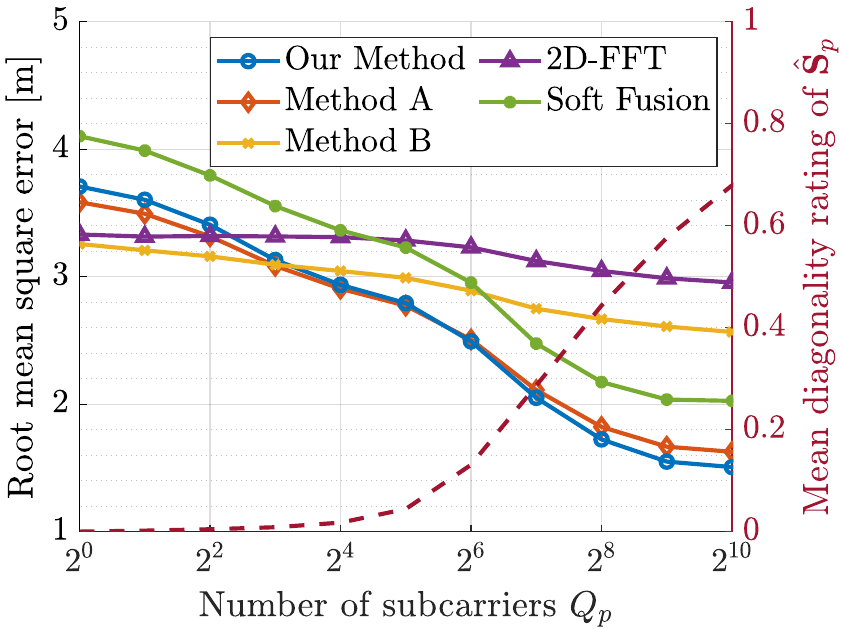}
\caption{Impact of the number of subcarriers when each radar pair as the same characteristics. The RMSE is compared for the different methods as a function of the number of subcarriers. The impact of the number of subcarriers on the diagonality of $\Hat{\mathbf{S}}_p$ is represented by the dashed line.}
\label{fig:results_Q}
\end{figure}

For each simulation, we analyze a fixed multistatic radar setup consisting of two STx and one SRx. This yields two radar pairs to locate three randomly placed targets. \figurename~\ref{fig:setup} shows the setup for one realization of the position of the targets and displays the combined log-likelihood map obtained with the method proposed in Section \ref{sec:MLF}. We evaluate the root mean square error (RMSE) for the presented methods through $15 000$ Monte-Carlo simulations.

Table~\ref{tab:RMSE} presents the results for different radar pair characteristics. It is observed that the proposed method outperforms all other compared techniques. When both radar pairs possess identical features, the proposed method proves to be more precise than the unweighted fusion of MUSIC outputs computed by method A. This highlights the significance of the signal power coefficient $\Hat{\overline{s}}_{p}(\boldsymbol\Phi_p,\boldsymbol\Theta_p)$. Nevertheless, it is crucial to consider the estimator employed to determine this coefficient as method B is not as accurate as the proposed method. Also, the improvement in RMSE is more significant when doubling the number of antennas than halving the noise variance of the estimated channel. 
%Additionally, the proposed method outperforms Soft Fusion, albeit with limited data transfer to the central processor. Therefore, when designing a system, the combination method should be selected based on the balance between accuracy and data transfer.

\balance In Section \ref{sec:ML_to_MUSIC}, we made the assumption that the matrix $\Hat{\mathbf{S}}_p$ is diagonal to ensure that the proposed method is equivalent to the complete $2K$-dimensional ML estimator. We now discuss the impact of this assumption. \figurename~\ref{fig:results_Q} displays the impact of the number of subcarriers $Q_p$ on the accuracy of the presented methods and evaluates its influence on the average diagonality of the matrix $\Hat{\mathbf{S}}_p$ using a criterion defined in \cite{2017290}. The matrix is considered perfectly diagonal when the criterion is equal to 1 and balanced when it is equal to 0. With an increase in the number of subcarriers, the accuracy of the proposed method, of method A, and of the soft fusion increases. This is due to the improved reliability of the MUSIC algorithm outputs when the sample covariance matrix $\mathbf{R}_p$ is well estimated. It can be observed that the proposed method remains reliable and more accurate than the other methods even when $\Hat{\mathbf{S}}_p$ is not perfectly diagonal. 
\section{Conclusion}
In this paper, we propose a novel data-level fusion method derived from the ML framework for a multistatic OFDM radar. The method exploits the relationship between the MUSIC and ML estimators to perform a joint AoD/AoA-based localization of $K$ targets. Further, we demonstrate that a weighted combination of MUSIC outputs from each radar pair of the multistatic system can provide an efficient approximation of the ML estimator. This approach reduces the complexity of the $2K$-dimensional search required by the ML estimator into $K$ two-dimensional problems solved by MUSIC. 

Unlike classical data-level fusion rule, the proposed combination avoids the transmission of the full raw data observations. Instead, each radar pair solely transmits its estimated channel covariance matrix to the central processor.  

The performance of the proposed combination is compared to other fusion methods and its benefit is evaluated by numerical simulations. The study examines various system parameters, including the number of antennas, the noise variance, and the number of subcarriers, to assess their impact on localization accuracy. The proposed methodology could be expanded in future works to take advantage of the range and Doppler estimations of each radar pair to enhance localization accuracy. The proposed framework can also be extended to any type of multistatic radar that provides noisy channel estimates.

%\begin{figure}
%\centering
%\includegraphics[width=0.8\columnwidth]{EuCAP}
%\caption{Example of a figure caption.}
%\label{fig:eucap}
%\end{figure}

%\begin{figure*}
%\centering
%\subfloat[Caption 1]{\includegraphics[width=0.5\columnwidth]%{EurAAP}%
%\label{fig:left figure}}
%\hfil
%\subfloat[Caption 2]{\includegraphics[width=0.5\columnwidth]%{EuCAP}%
%\label{fig:right figure}}
%\caption{Example of a double column floating figure using two subfigures.}
%\label{fig:double column}
%\end{figure*}
%
%\begin{table}
% % increase table row spacing, adjust to taste
%\renewcommand{\arraystretch}{1.3}
% if using array.sty, it might be a good idea to tweak the value of
% \extrarowheight as needed to properly center the text within the cells
%\caption{An Example of a Table}
%\label{table_example}
%\centering
% % Some packages, such as MDW tools, offer better commands for making tables
% % than the plain LaTeX2e tabular which is used here.
%\begin{tabular}{|c||c|}
%\hline
%One & Two\\
%\hline
%Three & Four\\
%\hline
%\end{tabular}
%\end{table}

\bibliographystyle{IEEEtran}
\bibliography{IEEEabrv,References}

\begin{thebibliography}{10}
\providecommand{\url}[1]{#1}
\csname url@samestyle\endcsname
\providecommand{\newblock}{\relax}
\providecommand{\bibinfo}[2]{#2}
\providecommand{\BIBentrySTDinterwordspacing}{\spaceskip=0pt\relax}
\providecommand{\BIBentryALTinterwordstretchfactor}{4}
\providecommand{\BIBentryALTinterwordspacing}{\spaceskip=\fontdimen2\font plus
\BIBentryALTinterwordstretchfactor\fontdimen3\font minus
  \fontdimen4\font\relax}
\providecommand{\BIBforeignlanguage}[2]{{%
\expandafter\ifx\csname l@#1\endcsname\relax
\typeout{** WARNING: IEEEtran.bst: No hyphenation pattern has been}%
\typeout{** loaded for the language `#1'. Using the pattern for}%
\typeout{** the default language instead.}%
\else
\language=\csname l@#1\endcsname
\fi
#2}}
\providecommand{\BIBdecl}{\relax}
\BIBdecl

\bibitem{9941042}
C.~Chen, H.~Song, Q.~Li, F.~Meneghello, F.~Restuccia, and C.~Cordeiro, ``Wi-fi
  sensing based on ieee 802.11bf,'' \emph{IEEE Communications Magazine},
  vol.~61, no.~1, pp. 121--127, 2023.

\bibitem{704504}
F.~Castanedo, ``A review of data fusion techniques,'' \emph{The Scientific
  World Journal}, pp. 1--19, 2013.

\bibitem{201328}
B.~Khaleghi, A.~Khamis, F.~Karray, and S.~Razavi, ``Multisensor data fusion: A
  review of the state-of-the-art,'' \emph{Information Fusion}, vol.~14, no.~1,
  pp. 28--44, 2013.

\bibitem{6994289}
H.~Jiang, J.-K. Zhang, and K.~M. Wong, ``Joint dod and doa estimation for
  bistatic mimo radar in unknown correlated noise,'' \emph{IEEE Transactions on
  Vehicular Technology}, vol.~64, no.~11, pp. 5113--5125, 2015.

\bibitem{8918315}
Y.~Li, X.~Wang, and Z.~Ding, ``Multi-target position and velocity estimation
  using ofdm communication signals,'' \emph{IEEE Transactions on
  Communications}, vol.~68, no.~2, pp. 1160--1174, 2020.

\bibitem{Falcone}
P.~Falcone, F.~Colone, A.~Macera, and P.~Lombardo, ``Two-dimensional location
  of moving targets within local areas using wifi-based multistatic passive
  radar,'' \emph{IET Radar, Sonar \& Navigation}, vol.~8, pp. 123--131,
  February 2014.

\bibitem{17564}
P.~Stoica and A.~Nehorai, ``Music, maximum likelihood, and cramer-rao bound,''
  \emph{IEEE Transactions on Acoustics, Speech, and Signal Processing},
  vol.~37, no.~5, pp. 720--741, 1989.

\bibitem{101049}
M.~Richards, J.~Scheer, and W.~Holm, \emph{Principles of Modern Radar}.\hskip
  1em plus 0.5em minus 0.4em\relax SciTech Pub., 2010, no. vol.~3.

\bibitem{1057561}
P.~Swerling, ``Probability of detection for fluctuating targets,'' \emph{IRE
  Transactions on Information Theory}, vol.~6, no.~2, pp. 269--308, 1960.

\bibitem{1164557}
M.~Wax and T.~Kailath, ``Detection of signals by information theoretic
  criteria,'' \emph{IEEE Transactions on Acoustics, Speech, and Signal
  Processing}, vol.~33, no.~2, pp. 387--392, 1985.

\bibitem{2017290}
K.~Alyani, M.~Congedo, and M.~Moakher, ``Diagonality measures of hermitian
  positive-definite matrices with application to the approximate joint
  diagonalization problem,'' \emph{Linear Algebra and its Applications}, vol.
  528, pp. 290--320, 2017.

\end{thebibliography}
\end{document}